\documentclass[a4paper,12pt,oneside,english]{report}
\usepackage{natbib} 
\usepackage{mathptmx}
\usepackage{amsmath}
\usepackage{amsfonts}
\usepackage{amssymb}
\usepackage{rotating}
\usepackage{subfigure}

\usepackage{graphicx}
\usepackage{babel}
\usepackage{epsfig}

\def\jgr{{\it Journal of Geophysical Research (Space Physics)}}
\def\grl{{\it Geophys. Res. Lett.,}}

\def\apjl{{\it Astrophys. Jou. Lett.}}
\def\aa{{\it Astron. \& Astrophys. Lett.}}
\def\aap{{\it Astron. \& Astrophys.}}

\def\asr{{\it Adv. Space Res.}}
\def\ssr{{\it Space Sci. Rev.}}
\def\solphys{{\it Sol. Phys.}}

\def\planss{{\it Planet. Space Sci.}}
\def\jastp{{\it JASTP.}}
\def\joaa{{\it Jou. of Astrophys. \& Astron.}}

\makeatother
\begin{document}

\Large{{\bf{Unique Observations of a Geomagnetic SI${^{+}}$ -- SI${^{-}}$ Pair: Solar 
Sources and Associated Solar Wind Fluctuations.}}}\\\\
{\large{R. G. Rastogi,  $^{1}$ 
\large{P. Janardhan,  $^{1}$
\large{K. Ahmed, $^{2}$
\large{A.C. Das, $^{1}$
\large{Susanta Kumar Bisoi, $^{1}$
}}\\\\
${^{1}}$ Physical Research Laboratory,Astronomy \& Astrophysics Division, 
Ahmedabad 380 009, India.\\
email: profrastogi@yahoo.com; jerry@prl.res.in; acd@prl.res.in; susanta@prl.res.in ~~ \\\\
${^{2}}$ Indian Institute of Geomagnetism, Navi Mumbai, 410218, India.\\
email:ahmedk@iigm.res.in
\section*{Abstract}

{\it{The paper describes the occurrence of a pair of oppositely directed sudden impulses (SI), 
in the geomagnetic field ($\Delta$X), at ground stations, called SI${^{+}}$ -- SI${^{-}}$ 
pairs, that occurred between 1835 UT and 2300 UT on 23 April 1998. The SI${^{+}}$ -- SI${^{-}}$ 
pair, was closely correlated with corresponding variations in the solar wind density, 
while solar wind velocity and the southward component of the interplanetary magnetic field 
(Bz) did not show any correspondence.  Further, this event had no source on the visible 
solar disk.  However, a rear-side partial halo coronal mass ejection (CME) and an 
M1.4 class solar flare behind the west limb, took place on 20 April 1998, the date 
corresponding to the traceback location of the solar wind flows.  This event presents 
empirical evidence, which to our knowledge, is the best convincing evidence for the 
association of specific solar events to the observations of an SI${^{+}}$ -- SI${^{-}}$ 
pair.  In addition, it shows that it is possible for a rear side solar flare to propagate 
a shock towards the earth.}}

\section*{Introduction}

It is well known that space weather events observed at 1 AU are all linked to the dynamic 
evolution of the solar photospheric magnetic field.  This evolution, in conjunction with 
solar rotation, drives space weather through the continuously changing conditions of 
the solar wind and the interplanetary magnetic field (IMF) within it.  In spite of the 
fact that there have been substantial observations and discussions on the close 
correspondence between solar wind parameters at 1 AU and ground based geomagnetic field 
variations \cite{Dun61, HMa87, GoL98, LuB98} it is not straightforward, under this broad 
framework, to pinpoint either the solar origins of specific space weather events or find 
specific correlations between solar wind parameters at 1 AU and ground based magnetic 
observations.  This is because such signatures are generally weak and are usually washed 
out or masked by the large variety of interactions that can take place both in the 
interplanetary medium and within the earth's magnetosphere.  Space weather events are 
however, often preceded by the arrival at 1 AU of strong interplanetary (IP) shocks. Since 
such storms can have adverse effects on human technologies, the study of IP shocks can 
yield important inputs for numerical models that simulate the propagation of solar-initiated 
IP disturbances out to 1 AU and beyond. 

On the other hand, solar sources of space weather events can range from coronal mass ejections 
(CME), very energetic solar flares, filament eruptions and corotating interaction regions (CIR).   
Though a vast majority of such events are caused by explosive and energetic solar events like CME's 
and flares, some recent studies have unambiguously associated large space weather events at 1 AU, 
like ``{\it{solar wind disappearance events}}", to small transient mid-latitude corona holes butting 
up against large active regions at central meridian \cite{ JaF05, JaF08, JDM08}.  These studies 
have provided the first observational link between the sun and space weather effects at 1 AU, 
arising entirely from non-explosive solar events.

Though the very first observations, by the Mariner 2 spacecraft in 1962, of interplanetary shock 
waves  showed the possibility of the existence of double shock ensembles in the interplanetary 
medium \cite{SoC64}, the existence of such shock pairs was firmly established only some years 
later, by the careful analysis of plasma and magnetic field measurements associated with shocks 
\cite{Bur70, LOB70}. However, the very unusual plasma and field variations associated with 
these structures prompted \cite{SCo65} to suggest that the first or `${\it{forward}}$' shock would 
give rise to a positive H-component at ground based observatories while the second or 
${\it{`reverse}}$' shock would cause an oppositely directed or negative change in the H-component 
of the earth's horizontal field, as measured along the local geomagnetic meridian (H).  These 
impulses, referred to in the rest of the paper as sudden impulse or SI${^{+}}$ -- SI${^{-}}$ pairs, 
were typically separated by a few hours in time and were hypothesized, as already stated, to be 
caused by the arrival at 1 AU of the forward and reverse shock pair convected towards the earth by 
the solar wind.  \cite{RCS65} described worldwide occurrences of such SI${^{+}}$ -- SI${^{-}}$ 
pairs and suggested that they were associated with solar disturbances driving interplanetary shocks 
at highly oblique angles to the solar wind streaming direction. They however, did not find any solar 
activity or associated occurrences of solar radio emission during the period of SI${^{+}}$ -- SI${^{-}}$ 
pairs.   In a more recent study of a number of SI${^{+}}$ -- SI${^{-}}$ pairs, covering the period 
1995--1999, \cite{TaA02} concluded that the observed SI${^{-}}$ (or negative impulses) in their 
sample were not associated with reverse shocks and showed no pre- 
%
%--------------------------begin figure 1 -----------------------------------
\begin{figure}[h]
\begin{center}
\includegraphics[width=0.56\textwidth,angle=0]{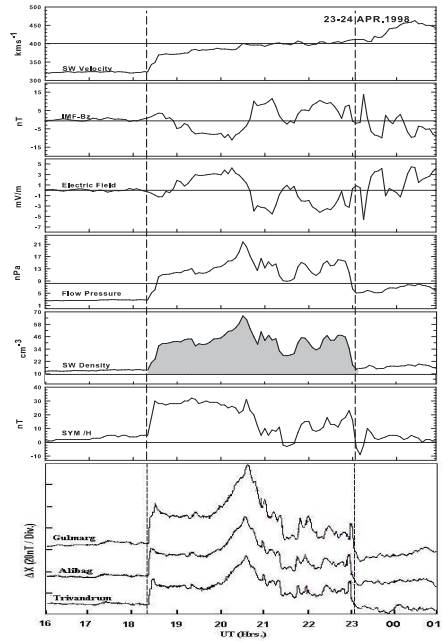}
\end{center}
\caption{The first six panels starting from the top show the variations as a function of 
time in UT on 23--24 April 1998 of the solar wind velocity, Bz, the Electric field, the solar 
wind flow pressure, the solar wind density (shaded grey) and the the symmetrical H Field 
respectively.  Observations of the H field at Indian geomagnetic stations Gulmarg, Alibag 
and Trivandrum are shown in the bottom most panel.  The pair of dashed vertically oriented 
parallel lines in all panels demarcate the times 1835 UT and 2300 UT.  These times 
correspond respectively to the times of the SI${^{+}}$ impulse and the SI${^{-}}$ impulse 
in the H field, that were observed at Indian stations.}
\label{swparam}
\end{figure}	     
%--------------------------end figure 1 -----------------------------------
%
\noindent ferential association to any particular kind of solar wind structure, like high and 
low speed stream interface discontinuities or front boundaries of interplanetary magnetic clouds.

Early theoretical support came from \cite{Dry70, Dry72} who introduced the physics of finite 
electrical conductivity within the forward and reverse shock pairs, to derive reasonable 
first order predictions for the observed distribution of solar wind speed, density, and 
temperature and this was followed up by several other early papers\cite{EDr70, SDr72, DrS75} 
on similar lines.  In more recent times, there have been a number of theoretical models that 
have used inputs from solar data to predict the arrival of IP shocks and IP CMEs at earth 
\cite{Ods03, VaW05, ToS05, DeS06}, including the well known Hakamada-Akasofu-Fry model (HAFV2), 
\cite{FrD03} which is the only model, to date, to have been substantially validated in an 
operational forecasting environment \cite{SmD09, SmS09} during solar cycle 23. 

\section*{The SI${^{+}}$ -- SI${^{-}}$ Pair of 23 April 1998}

An  SI${^{+}}$ -- SI${^{-}}$ pair was identified at three Indian geomagnetic observatories 
on 23-24 April 1998. Figure {\ref{swparam}} (bottom-most panel) shows the tracings of H 
magnetograms (projected onto the X or geographic north direction and marked $\Delta$X in Fig. 
\ref{swparam}) on 23-24 April 1998 at the three Indian stations Gulmarg, Alibag and Trivandrum 
respectively.  A sudden positive impulse in H was recorded at all three Indian observatories 
at 1835 UT (23 Apr. 2335 LT) followed by a sudden negative impulse at 2300 UT (24 Apr. 0430 
LT). During the time interval between the SI${^{+}}$ impulse and the SI${^{-}}$ impulse, the 
amplitude of H first decreased and then attained a peak of 44 nT at Trivandrum that 
progressively increased to 54 nT at Alibag and 76 nT at Gulmarg. Between 2100-2300 UT large 
fluctuations were recorded at all stations. The fluctuation in H, at all stations were remarkably 
similar to each other with the amplitude increasing from Trivandrum to Gulmarg. Also shown 
in Fig.\ref{swparam} (starting from the top and going down) are the corresponding variations 
of the solar wind velocity, the IMF-Bz, the interplanetary electric field, the solar wind flow 
pressure, the solar wind density and the symmetrical H Field respectively. The curve for 
the solar wind density, as observed by the Advanced Composition Explorer (ACE; \cite{StF98}) 
has been shaded grey, in Fig.\ref{swparam}, in the region between the SI${^{+}}$ -- SI${^{-}}$ 
pair.  The vertically oriented dashed parallel lines in all panels demarcates the time interval 
between the SI${^{+}}$ impulse and the SI${^{-}}$ impulse.  It is to be noted that the symmetric H 
field (SYM/H), characterizing the mean variation of H at all middle latitude stations around 
the world, too had remarkably similar variations as the H at Indian stations. This therefore 
implies that the SI${^{+}}$ -- SI${^{-}}$ pair was a global event. 

\cite{Rpa75} had shown that solar plasma moving towards the earth's magnetosphere with
the velocity, V and having a frozen-in magnetic field normal to the ecliptic (IMF-Bz) is 
equivalent to an electric field, E${_{sw}}$ = (-V $\times$ Bz), which is transmitted without 
any time delay to the polar region and then to the low latitude ionosphere. This belongs to a 
process known as overshielding electric field in the magnetosphere which has been 
extensively studied \cite{Nis68, Vas70, SWF88, WoG01, GoS02}.  Prompt penetration occurs 
due to the slow response of the shielding electric field at the inner edge of the ring 
current that opposes the time varying convection field, in the presence of an IMF-Bz.  
The time scale of this process is generally of the order of an hour but can sometimes 
be longer \cite{Vas70,SBl84}.  During a period of sudden northward turning of the IMF, 
from a steady southward configuration, the convective electric field shrinks while the 
shielding electric field takes a longer time to decay and produce a residual electric 
field, known as the overshielding electric field. The direction of this field is opposite 
to the normal direction of the ionospheric electric field.  The E${_{sw}}$ has a direction 
of dusk-to-dawn for positive IMF-Bz and dawn-to-dusk for negative IMF-Bz. 
%
%--------------------------begin figure 2 -----------------------------------
\begin{figure}[ht]
\begin{center}
\includegraphics[width=0.4\textwidth,angle=0]{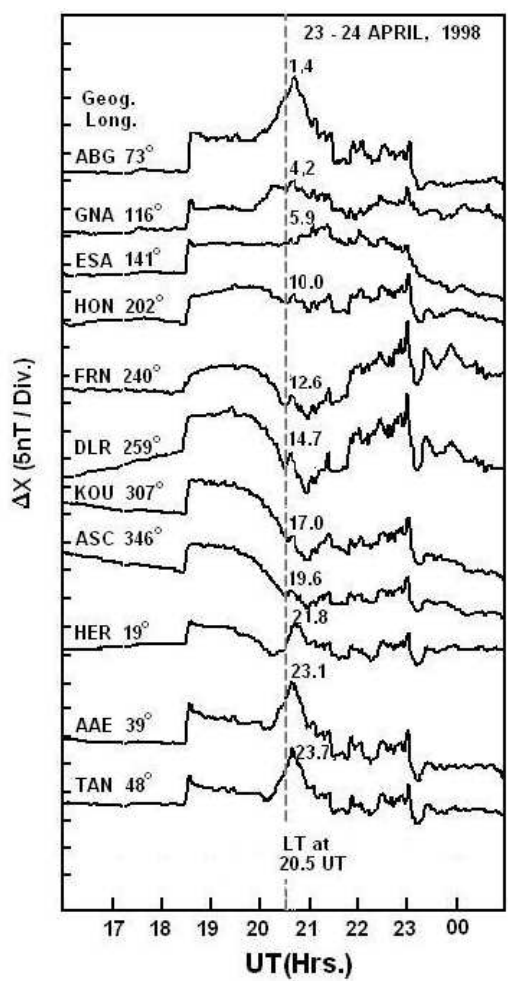}
\end{center}
\caption{Variation of the horizontal component of the geomagnetic field projected 
on to the x-direction at eleven low latitude stations around the world on 23 -- 24 
April 1998.  The vertical dashed line is marked at 20.5 UT and shown alongside it, 
for each curve, is the corresponding local time at each station.  Also indicated 
on the left of each curve is the geographic longitude of the station.} 
\label{deltax}
\end{figure}	      
%--------------------------end figure 2 -----------------------------------
%

It can be seen from Fig. \ref{swparam} that $\Delta$X at Indian stations just 
after the SI${^{+}}$ at 1835 UT (around local midnight) had gradually decreased till 2000 UT. 
This effect is due to the prompt penetration of the electric field when the IMF-Bz is negative 
during the period. At around 2000 UT, $\Delta$X again increased suddenly to values much above 
the first impulse level. Correspondingly, it can be seen that IMF-Bz turned from southward 
to northward at this instant implying that it is due to the overshielding condition 
described above.  After this, the level of $\Delta$X went down, with some oscillations and 
finally came down to normal level at 2300 UT. It is interesting to note that 
the fluctuations in $\Delta$X between 2130-2200 UT were very well correlated with 
the solar wind density rather than with the IMF-Bz or the solar wind speed.  The 
SI${^{+}}$ at 1835 UT  was associated with a sudden increase of both the solar wind 
density and speed causing a sudden pressure on the magnetosphere (as first suggested 
by \cite{Gol59}). The SI${^{-}}$ at 2300 UT was associated with the sudden decrease 
of solar wind density. 

\subsection*{Global Geomagnetic Fields}

Figure \ref{deltax} shows the variation of $\Delta$X from eleven low latitude stations around 
the world on 23-24 April 1998.  The stations, starting from Alibag (ABG -- Lat. 18.64; Long.72.87), 
(uppermost curve) and arranged in increasing order of geographic longitude are respectively, 
Gnangara (GNA -- Lat. -31.78; Long. 115.95), Esashi (ESA -- Lat. 39.24; Long. 141.35), Honululu (HON 
-- Lat. 21.32; Long. 202.00), Frenso (FRN -- 37.10; Long. 240.30), Del Rio (DLR -- Lat. 29.49; Long. 
259.08), Kourou (KOU -- Lat. 2.21; Long. 307.27), Ascension Island (ASC -- Lat. -7.95; Long. 345.62), 
Hermanus (HER -- Lat. -34.42; Long. 19.23), Addis Ababa (AAE -- Lat. 9.02; 38.77) and 
Tanananarive (TAN -- Lat. -18.92; Long. 47.55).  The geographic longitude of the ground stations 
are indicated at the left of each curve in Fig. \ref{deltax}. Also indicated to the right of the 
vertical dashed line (marked at 20.5 Hrs UT in Fig. \ref{deltax}) for each curve, is the local time at 
20.5 hrs UT.  The stations chosen range from geographic longitudes of 19${^{\circ}}$ to 
346${^{\circ}}$ corresponding to local times of $\sim$22 hr. through the midnight, dawn, noon to 
dusk (20 hr). 

The negative IMF-Bz between 1835 -- 2000 UT caused a decrease of $\Delta$X at stations in 
the night sectors (HER, AAE, and TAN ) and an increase at stations in the mid day sector 
(FRN, DLR, KOU and ASC).  Around 20.50 UT, $\Delta$X showed strong positive peaks at AAE, TAN 
and ABG, no change at ESA and HON and negative peaks at FRN, DLR and KOU.  This data conforms 
very well with the process of prompt penetration and overshielding electric field 
\cite{Nis68, Vas70, Rpa75, SWF88} wherein, a southward IMF-Bz (between 1835 and 2000 UT) 
would cause a decrease of $\Delta$X at night side stations and a increase of $\Delta$X at day 
side stations of the earth while a northward turning of IMF-Bz would produce a strong
\noindent  positive $\Delta$X at stations in the night sector and negative $\Delta$X at stations 
in the day side sector due to the imposition of either a dusk-to-dawn or dawn-to-dusk electric 
field.  The fluctuations in $\Delta$X between 2130 - 2300 UT are synchronous at all stations 
in the day as well as in night sectors, suggesting the effect to be due to solar wind 
flow pressure and not due to the IMF-Bz.  It is important to note here that the solar wind 
density fluctuations virtually mirrors those seen in $\Delta$X at ground stations, thereby 
implying that the solar wind density was the main key or driver for this event.  Qualitatively,
the fluctuations seem to be independent of the latitude.  The dominant parameter is the 
solar wind pressure that makes the magnetosphere shrink and expand self similarly, with 
some scaling factor depending on the pressure.  This is reflected in the magnetic field data 
at all latitudes and longitudes in a configuration where the IMF appears to have no role to 
play. Thus, this was a unique space weather event in which one could unambiguously associate 
solar wind density variations with variations in $\Delta$X at ground stations while no 
such changes were seen in the solar wind speed or magnetic field.

\section*{Solar Source Locations}

It is well known that due to the rotation of the sun ($\Omega$~=~1.642$\times$10${^{-4}}$ deg s${^{-1}}$), a radially directed outflow of solar wind from the sun will trace out an Archimedean spiral through the interplanetary medium.  For a steady state solar wind with a velocity of 430 km s${^{-1}}$, the tangent to this spiral, at 1 AU, will make an angle of 45${^{\circ}}$ with the 
radial vector from the Sun \cite{Sch90}.  As a consequence, the longitudinal offset 
($\phi$${_{R}}$) of a solar wind stream with a velocity v, when traced backwards from a 
distance R${_{1}}$ (say 1 AU) to a distance R${_{2}}$, will 
be $\phi$${_{R}}$~=~$\Omega$(R${_{1}}$~-~R${_{2}}$)/v.  We can thus project the observed solar wind 
velocities back to the sun to determine the sources of the solar wind flows at the sun.  The 
earliest instance of using such a technique to trace solar wind outflows back to the sun was by 
\cite{Ric75}.  For the present event, we have back-projected the observed ACE velocities along 
Archimedean spirals to the source surface at 2.5 R${_{\odot}}$ to determine its solar source 
location.  Though this method is generally applicable to a steady-state flow of the solar wind, it 
has also been applied in cases when the solar wind outflows were not steady-state and highly 
non-radial.  For example, during the well know disappearance event of 11 May 1999, the work by 
\cite{JaF05, Jan06} has shown that solar source locations determined by the traceback technique, 
using constant velocities along Archimedean spirals, do not have significant errors even though 
the solar wind flows were known to be highly non-radial during that period.  In the case under 
discussion the solar wind flows would have been highly kinked and non-radial due to the 
propagating forward and reverse shocks arising from the optically occulted flare and the rear 
side CME.  Therefore, if the SI${^{+}}$ -- SI${^{-}}$ pair had a source on the solar disk, 
the ambiguity about the location of the source region would be within reasonable errors as shown by 
\cite{JaF05, Jan06}. Figure {\ref{armap}} shows a map of the solar photosphere indicating the 
locations of the active regions.  The back projected region of
%
%--------------------------begin figure 3 -----------------------------------
\begin{figure}[ht]
\begin{center}
\includegraphics[width=0.45\textwidth,angle=0]{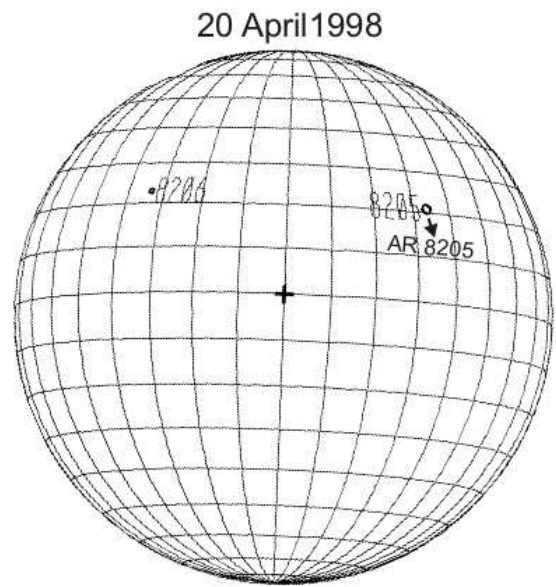}
\end{center}
\caption{Map of the solar photosphere on 20 April 1998 corresponding to the 
back projected region of the solar wind flows obsrved at 1 AU.  The map shows 
the locations of the large active regions with active region AR8205 indicated 
by an arrow for convenience.}
\label{armap}
\end{figure}
%--------------------------end figure 3 -----------------------------------
%
\noindent the solar wind flows go back to the vicinity of the large active region AR8205 
located at N21W25, to the west of central meridian on 20 April 1998, and indicated by a 
solid arrow in Fig. {\ref{armap}}.  Typically, the solar disk shows a large number of 
active regions during the rising phase of the solar cycle.  A detailed 
theoretical study by \cite{SDe03},has shown that solar wind outflows from active 
regions comprise $\leq$10\% during solar minimum and up to 30--50\% during solar 
maximum. However, the visible solar disk on 20 April 1998 showed no activity in 
terms of flares or CME's and had only two active regions AR8206 and AR8205 as shown 
in Figure \ref{armap}.  The Active region AR8206 was smaller than AR8205 being around 
245 millionth of the solar disk in size as compared to 312 millionth of the solar disk 
for AR8205.  Also, AR8205 was less than 30${^{\circ}}$ west of central meridian as 
compared to over 40${^{\circ}}$ east of central meridian  for AR8206.  It may be noted 
that a central meridian location would imply that any activity like a large CME or 
flare would be earth directed. However, there was no flare or CME on the entire visible 
solar disk on 20 April. Images from the Extreme-ultraviolet Imaging Telescope 
(EIT; \cite{DeA95}) and the Michelson Doppler Imager (MDI, \cite{ScB95}) onboard the 
Solar and Heliospheric Observatory (SoHO;\cite{DFP95}) were also examined carefully 
to confirm that there was no other possible source regions on the solar disk on 
20 April 1998. 
 
\subsection*{The Rear side CME and Optically Occulted Flare of 20 April 1998}

On 20 April 1998, a rear side, fast ($\sim$1850 km s${^{-1}}$) partial halo CME occurred in 
association with an optically occulted GOES M1.4 class flare which took place just behind 
the limb at S43W90.  It must be pointed out here that most forecasters of space weather events 
generally ignore the possibility that a limb or backside solar explosive event could propagate 
a disturbance towards the earth. However, there have been some instances where such cases have been 
studied and reported in recent times \cite{McD06, SmS09, SmD09}.  

The GOES M1.4 flare at S43W90 was first detected in 1-8 \AA ~band at 09:15 UT on 20 April 1998 
and reached its maximum at 10:21 UT.  The rear side partial halo CME was first seen in the LASCO 
coronograph C2, at 10:04:51 UT on 20 April 1998, as a bright, sharply defined loop structure 
spanning $\sim$80${^{\circ}}$ in latitude and extending to $\sim$3.1 R${_{\odot}}$. The same 
was first observed by C3 at 10:45:22 UT.  Both the CME and the flare have been extensively 
studied and reported \cite{BaP01, Sim00, Sim02} and it has been shown that the CME, which was 
radio loud \cite{Gop00}, actually pushed aside pre-existing streamers while moving beyond the 
LASCO C3 field of view.  Since this was a rear-side CME the shock front that it drove would 
have been in a direction away from the earth.  In a study of the arrival time of flare driven 
shocks at 1 AU and beyond \cite{SSh85, JaB96} it has been assumed that the trailing edges of 
flare driven shock waves travel at roughly half the velocity of the shock in the flare radial 
direction.  It is therefore not unreasonable to assume that the trailing edges of the CME driven 
reverse shock would be much slower and could be convected outwards towards the earth by the 
solar wind.  The flare and the rear-side CME would thus provide the forward and reverse shocks 
to cause the SI${^{+}}$ and SI${^{-}}$ pair.  Figure {\ref{tot-mag}} shows hourly averaged value 
of the absolute magnitude of the magnetic field, as observed by the ACE spacecraft, as a function 
of time in UT.  It is expected that the strength of the magnetic field would increase at the 
forward shock or SI${^{+}}$ impulse and decrease at the reverse shock or SI${^{-}}$ impulse.  
It can be easily seen from Fig. {\ref{tot-mag}} that there is a sharp increase in the magnetic 
field at around 1835 UT, corresponding to the arrival of the forward shock associated with the 
SI${^{+}}$ impulse and a decrease in the magnetic field at 2300 UT corresponding to the arrival 
of the reverse shock associated with the SI${^{-}}$ impulse. The vertically oriented dashed 
parallel lines in Fig. {\ref{tot-mag}} are marked at 1835 UT and 2300 UT, the time corresponding 
to the  SI${^{+}}$ and SI${^{-}}$ impulse respectively.
%--------------------------begin figure 4 -----------------------------------
\begin{figure}[ht]
\begin{center}
\includegraphics[width=0.5\textwidth,angle=0]{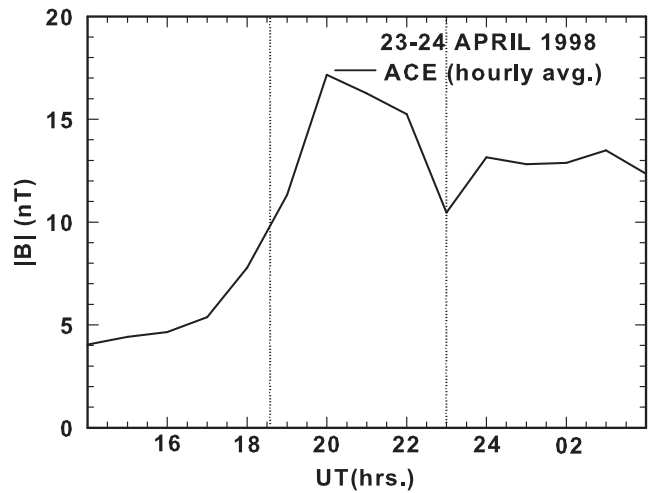}
\end{center}
\caption{Hourly averaged total magnetic field as a function of time in UT as observed by the 
ACE spacecraft located at the L1 Lagrangian point at 1 AU.  The vertically oriented 
dashed parallel lines at 1835 UT and 2300 UT correspond to the time of the SI${^{+}}$ and 
SI${^{-}}$ impulse respectively.} 
\label{tot-mag}
\end{figure}
%--------------------------end figure 4 -----------------------------------
%
\section*{Discussion and Conclusions}

From an observational point of view, the present work has been able to link interplanetary 
structure during this particular event with worldwide magnetospheric response, using the 
Indian magnetic observatories to provide the first clue.  In particular, this event has been 
unique in that the solar wind density variations have played a key role, as seen through the 
close correspondence between the fluctuations in the solar wind densities and the $\Delta$X at 
ground stations while no such changes were seen in the solar wind speed or magnetic field. 
Though there has been a large body of work, over the past four decades, that have addressed 
the issues concerned with forward/reverse shock pairs, their manifestation at 1 AU and their 
relation to specific solar events, we believe that this paper presents empirical 
evidence, which to our knowledge, is the best convincing evidence for the association of specific 
solar events to the observations of an SI${^{+}}$--SI${^{-}}$ pair. In addition, it shows 
that it is possible for a rear side solar event to propagate a shock towards the earth.
 
We have seen that the SI${^{+}}$ impulse 1835 UT  was associated with a sudden 
increase of both the solar wind density and speed causing a sudden pressure on the 
magnetosphere while the SI${^{-}}$ at 2300 UT was associated with the sudden decrease of 
solar wind density. The southward IMF-Bz between 1835-2000 UT caused a decrease of 
$\Delta$X at night side stations and a increase of $\Delta$X at day side stations of 
the earth due to the imposition of a sudden electric field caused by the prompt penetration 
of electric field to low latitudes.  As stated earlier, a northward turning of IMF-Bz produces 
a strong positive $\Delta$X at stations in a night sector and negative $\Delta$X at stations 
in the day sight sector due to the effect of overshield electric field which is in a direction 
opposite to the normal electric field in the ionosphere. Between 2015-2300 UT the fluctuations 
in $\Delta$X were similar at all stations in the day or night sectors and were 
well correlated with the fluctuations in solar wind flow pressure, reflecting the shrinking 
and expansion of the magnetopause as a result of strong solar wind pressure variation. 
 
The solar event lasting only for some 4 to 5 hours showed signatures of all mechanisms 
involving solar -- magnetosphere -- ionosphere relationships. The arrival at 1 AU of the 
forward and reverse shock pair associated with the SI${^{+}}$ and SI${^{-}}$ respectively is 
clearly seen in the behaviour of the hourly averaged values of the total magnetic field, which 
shows a sharp increase at $\sim$1835UT and a decrease at $\sim$2300 UT.  The effect of sudden 
changes in the solar flow pressure due to a change of only the solar wind density have been 
clearly identified. The effect of the slowly varying IMF-Bz has been shown to impose dusk-to-dawn 
or dawn-to-dusk electric field globally, depending on the southward or northward 
turning of the IMF-Bz. Though theoretical first order predictions for the observed distribution 
of solar wind speed, density, and temperature (as in Fig.1) during the propagation of forward 
and reverse shock pairs were derived four decades ago \cite{Dry70, Dry72}, the analysis of 
the event has been rewarding due to the relative quiet solar conditions prevailing as it 
allowed us to identify specific solar sources as the possible drivers of the SI${^{+}}$ and 
SI${^{-}}$ pair.  The only activity on Sun was the rear side CME and the associated solar flare. 
This is thus a very unique observation wherein, a pair of SI events have been shown to be 
associated with corresponding changes in the solar wind density while no such changes are seen 
in the solar wind speed or magnetic field.  Many more such events need to be observed, retrieved 
and studied, both from archival records and future observations, before a clearer understanding 
of the exact nature and physics behind such events is obtained.  High resolution, high dynamic 
range radio imaging techniques \cite{MeP06} can also provide useful information in this regard. 

\section*{Acknowledgments}

{\it{The authors would like to thank the two referees for their very constructive suggestions that 
have gone a long way in improving this paper.  One of the authors (JP) would like to thank Dr. 
Murray Dryer, for specific and focused comments.  The authors thank King, J.H. and Papatashvilli, 
N. of Adnet Systems, NASA, GSFC, the CDAWeb team, the ACE SWEPAM instrument team, the ACE Science 
Center and the EIT and MDI consortia for making data available in the public domain via the 
world wide web. SoHO is a project of international collaboration between ESA and NASA.}}

\end{document}